\documentclass[apl,reprint,showpacs,twocolumn,superscriptaddress]{revtex4-1}
\pdfoutput=1

\usepackage[colorlinks,linkcolor=blue,citecolor=blue]{hyperref}
\usepackage{graphicx}
\usepackage{natbib}
\usepackage{comment}
\usepackage{amsmath}
\usepackage{amssymb}
\usepackage{amsfonts}
\usepackage{upgreek}
\usepackage{gensymb}
\usepackage{textcomp}

\begin{document}


\title{Balanced double-loop mesoscopic interferometer based on Josephson proximity nanojunctions}

\author{Alberto Ronzani}
\email[]{alberto.ronzani@nano.cnr.it}
\affiliation{NEST, Istituto Nanoscienze-CNR and Scuola Normale Superiore,
             I-56127 Pisa, Italy}

\author{Carles Altimiras}
\affiliation{NEST, Istituto Nanoscienze-CNR and Scuola Normale Superiore,
             I-56127 Pisa, Italy}

\author{Francesco Giazotto}
\affiliation{NEST, Istituto Nanoscienze-CNR and Scuola Normale Superiore,
             I-56127 Pisa, Italy}


\begin{abstract}
We report on the fabrication and characterization of a two-terminal mesoscopic interferometer
based on three V/Cu/V Josephson junctions having nanoscale cross-section.
The junctions have been arranged in a double-ring geometry realized by metallic thin film 
deposition through a suspended mask defined by electron beam lithography.
Although a significant amount of asymmetry between the critical current of each junction is observed
we show that the interferometer is able to suppress the supercurrent to a level lower than 6 parts per thousand, 
being here limited by measurement resolution. The present nano-device is suitable for low-temperature
magnetometric and gradiometric measurements over the micrometric scale.
\end{abstract}


\maketitle

Transport properties of hybrid superconductor-normal metal structures at the mesoscopic scale
are understood in terms of the \textit{proximity effect}\cite{Taddei2005,Pannetier2000,Belzig1999,McMillan1968,McMillan1968a,Giazotto2010},
 which consists in the modification of the
electronic properties of a normal metal in clean contact with a superconductor.
Dissipationless transport can be established in superconductor-normal metal-superconductor
(SNS) junctions owing to the phase-dependent weak coupling across the diffusive normal metal
channel, which can persist over micrometric length scale. The current to phase relationship
is sinusoidal in the limit of weak links much longer than the superconducting coherence 
length $\xi_0$~\footnote{$\xi_0 = \sqrt{\hbar D / \Delta}$, where $D$ is the diffusion coefficient in the normal metal 
and $\Delta$ is the superconducting energy gap at the leads.}, so that SNS weak links are functionally equivalent
to superconductor-insulator-superconductor (SIS) Josephson junctions\cite{Josephson1962}, featuring however negligible capacitance
and much lower typical values of normal-state resistance\cite{Likharev1979}.

Several experimental works concerning superconducting interferometers based on SNS technology have been published
recently\cite{Wolbing2013,Ronzani2013,Nagel2011a,Angers2008a}. 
One commonly encountered practical issue is that the fabrication protocols do not allow to fully control
the quality of the SN interfaces. This results in non-negligible asymmetries in the magnitude of supercurrents between
nominally-identical SNS junctions, therefore severely limiting the sensitivity of the interferometer.
Here we propose the adoption of a double-loop SNS interferometer configuration
as a means to compensate for such asymmetries. This geometry 
has been proved successful in SIS systems in balancing devices intended for metrological\cite{Kemppinen2008},
quantum computation\cite{Chiarello2008} as well as sensing\cite{Dhong1983,Sharafiev2012} applications.
Yet, it has been proposed to fully suppress phase-coherent heat currents in caloritronic devices\cite{Martinez-Perez2013}.
With our approach we demonstrate the feasibility of balancing a mesoscopic SNS interferometer
to obtain supercurrent suppression ratio values well below one percent, making it an attractive
device element for superconducting quantum circuitry and nanoscale sensing applications.

\begin{figure}[t]
\includegraphics[width=\columnwidth]{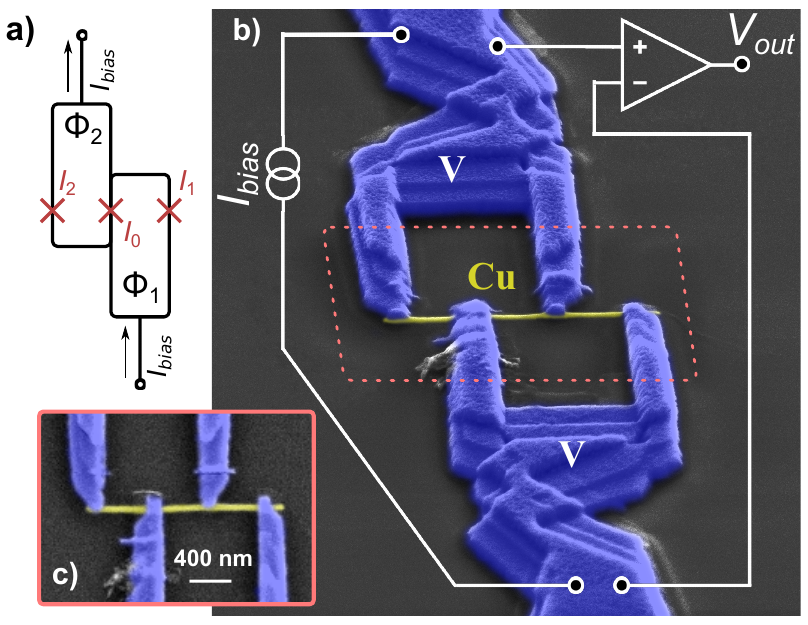}
\caption{
\textbf{a)} Functional scheme for a double-loop superconducting interferometer: $\Phi_{1,2}$ represent magnetic fluxes
linked to each loop; the critical current values for the three Josephson junctions are
labeled $I_{0,1,2}$.
\textbf{b)} Tilted scanning electron micrograph showing the interferometer in pseudocolors.
Copper (yellow) nanowire is $25\, \mathrm{nm}$ thick; the vanadium (blue) electrodes contacting the copper element
are $80\, \mathrm{nm}$ thick near the nanowire and $160\, \mathrm{nm}$ thick farther away on the loops as well as on the two terminals of
the device. Each loop spans a surface $\approx 1.18\, \upmu\mathrm{m}^2$. The standard setup scheme for
a four-wire measurement is overlaid on the micrograph.
\textbf{c)} Scanning electron micrograph showing the top view of the metallic nanowire. The interelectrode
spacing is approximately equal to $450\, \mathrm{nm}$, the nanowire is $45\, \mathrm{nm}$ wide.
\label{fig:sem}}
\end{figure}
The geometry of a double-loop SNS interferometer consists in the parallel
circuit of three Josephson junctions; this configuration defines two superconducting rings,
each of which is coupled to a separate magnetic flux \lbrack Fig.~\ref{fig:sem}a)\rbrack .
Figure~\ref{fig:sem}b) shows a scanning electron micrograph of our implementation of this type of interferometer, fabricated by standard
electron beam lithography on a suspended bilayer resist mask\cite{Dolan1977} 
 (1000~nm copolymer / 100~nm polymethyl-metacrylate) on top of an oxidized
silicon substrate. A 5~nm-thick adhesivant aluminium layer has been deposited at $-40^\circ$ via electron-beam evaporation 
in ultra-high vacuum conditions ($\simeq 10^{-9}\, \mathrm{Torr}$),
followed by 25~nm of copper at normal incidence; finally, the superconducting
leads have been realized by evaporating vanadium in two steps at opposing angles~($\pm 17^\circ$), 80~nm per step. 
Excess metal and resist have been removed by lift-off, then the sample has been inspected by scanning electron microscopy
and finally wire-bonded to a ceramic dual-in-line chip carrier.
The fabricated interferometer features three weak links \lbrack Fig.~\ref{fig:sem}c)\rbrack~consisting 
of a diffusive normal-metal wire having width and thickness of
45 and 25~nm, respectively; the inter-electrode spacing between vanadium leads is approximately equal to 450~nm.
Since the transverse extent of the copper wire is less than the superconducting coherence length $\xi_0$, 
at each vanadium electrode the 
local\footnote{That is, within distance comparable to $\xi_0$ from the interface with the superconductor.}
electronic density of states in the normal metal is expected to have a minigap close to the superconducting energy gap in the electrode itself,
so that the system can be pictured as having three independent weak links\cite{Pfeffer2013}.

\begin{figure}[t]
\includegraphics[width=\columnwidth]{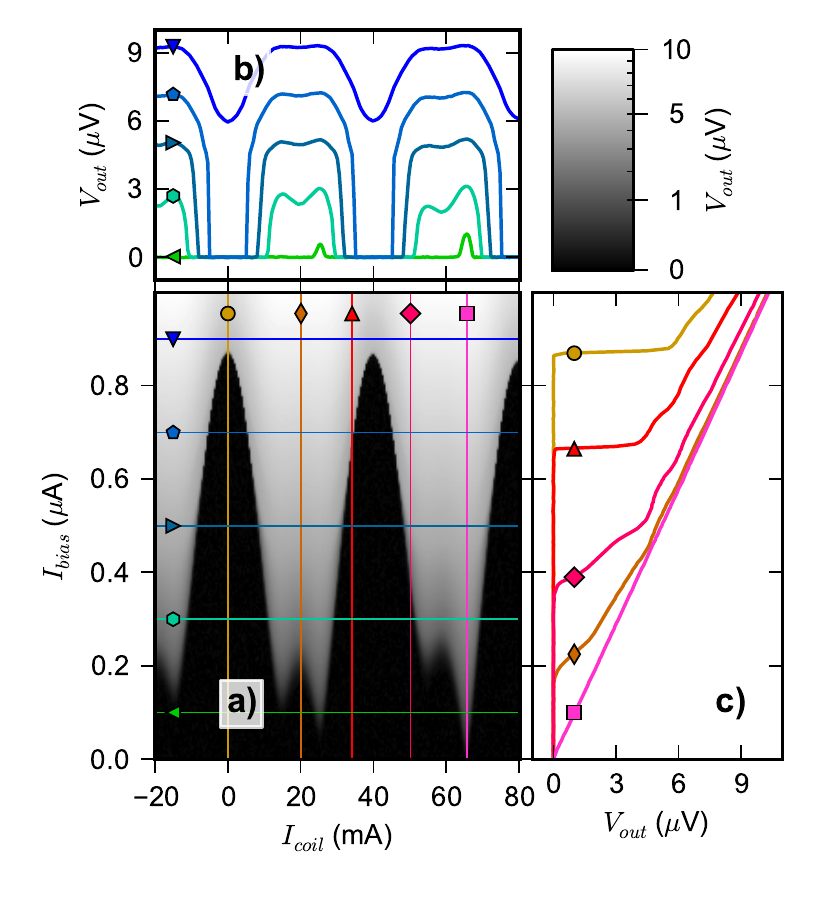}
\caption{
\textbf{a)} Greyscale map of the voltage response $V_{out}$ of the device at $240\, \mathrm{mK}$ as a function of both the biasing current
$I_{bias}$ and the current $I_{coil}$ feeding into the magnetic bias coil, which is proportional to the applied magnetic flux.
The grey level has been made proportional to the square root of $V_{out}$ in order to enhance the visual contrast of the switching points.
\textbf{b)} Flux to voltage characteristic curves extracted from $V_{out}(I_{bias}, I_{coil})$ data for fixed values of $I_{bias}$ marked 
as horizontal lines in panel \textbf{a} ($I_{bias} = 0.1, \, 0.3, \, 0.5, \, 0.7, \, 0.9 \, \mathrm{\mu A}$).
\textbf{c)} Current to voltage characteristic curves extracted from $V_{out}(I_{bias}, I_{coil})$ data for fixed values of $I_{coil}$ marked 
as vertical lines in panel \textbf{a} ($I_{coil} = 0, \, 20, \, 34, \, 50, \, 66 \, \mathrm{mA}$).
\label{fig:ivs}}
\end{figure}

The electron transport properties of the interferometer have been characterized in a filtered ${}^3\mathrm{He}$ cryostat 
down to $T \approx 240\,\mathrm{mK}$. 
Current vs voltage curves have been recorded by measuring the response to a
DC current bias chopped at a reference frequency ($f = 17\,\mathrm{Hz}$) using a lock-in amplifier (NF~Corp. model LI-5640). 
The voltage response of the interferometer has been characterized as a function of the current $I_{coil}$ feeding into the 
superconducting coil used to generate a magnetic field orthogonally to the
substrate of the sample, and for several temperatures in the range $0.24\mathrm{K} - 1.5\mathrm{K}$. Figure~\ref{fig:ivs}a) shows
the characterization of the interferometer at base temperature ($T\approx0.24\mathrm{K}$); the typical input referred voltage noise
for this setup has been measured to be $< 10\,\mathrm{nV_{rms}}$.
The device shows a remarkably linear voltage response for $I_{coil} = 66\, \mathrm{mA}$, the characteristic marked with a pink square in 
Fig.~\ref{fig:ivs}c), corresponding to a measured resistance $R_n \approx 10.3\,\Omega$.
Additionally, not exactly periodic modulation can be appreciated from the voltage vs flux characteristics shown in Fig.~\ref{fig:ivs}b),
indicating a slight asymmetry in the effective areas of the two superconducting loops.

The behavior displayed by the device can be understood in terms of a compact model:
assuming a sinusoidal current-phase relationship for each Josephson junction 
and imposing flux quantization\cite{Doll1961, Deaver1961}
 constraints one obtains for the total supercurrent $I_{SC}$ flowing through the interferometer
\begin{equation}
	\label{eqn:model0}
	\left\{
	\begin{aligned}
	I_{SC} & = I_0 \sin(\delta_0) + I_1 \sin(\delta_1) + I_2 \sin(\delta_2) \\
	\delta_1 & = \delta_0 + 2 \pi \Phi_1/\Phi_0 \\
	\delta_2 & = \delta_0 - 2 \pi \Phi_2/\Phi_0 \, ,
\end{aligned} 
\right.
\end{equation}
where $\Phi_0 = h/(2e)$ is the flux quantum; $I_i$ and $\delta_i$ respectively represent the critical current and phase 
difference values for the three Josephson junctions, $i = 0, 1, 2$ referring to the central, left and right weak link.

Within this model, the superconducting interferometer is able to conduct a dissipationless current whose maximum value $I_c$
is a function of the magnetic flux values $\Phi_1$ and $\Phi_2$ linked to each superconducting loop.
The value of $I_c$ for fixed flux biasing is thus
\begin{equation}
	\begin{split}
	I_c(\Phi_1, \Phi_2)  = \max_{\delta_0} \, [ \, &I_0 \sin(\delta_0) + \\
	                     + &I_1 \sin(\delta_0 + 2 \pi \Phi_1/\Phi_0) \\
			     + &I_2 \sin(\delta_0- 2\pi \Phi_2/\Phi_0) \, ] \, .
	\label{eqn:modelic}
	\end{split}
\end{equation}

A key point is that the critical current of the interferometer is given by the magnitude of the \emph{vector sum} of the 
critical currents of the three Josephson junctions, with $2 \pi \Phi_1/\Phi_0$ and $2 \pi \Phi_2/\Phi_0$ playing the role 
of angular displacements between the current vectors representing the lateral junctions with respect to the central one.
As such, one can in principle achieve perfect critical current suppression at appropriate $\Phi_1, \, \Phi_2$ values
as long as $I_0,\,I_1,\,I_2$ satisfy the triangle inequality:
\begin{equation}
	\left\{ \, \,
	\begin{aligned}
		\left| r_1 - r_2 \right| &\le& 1 \\
		 r_1 + r_2 \, &\ge& 1
	\end{aligned}
	\right. \, \, ,
	\label{eqn:triangle}
\end{equation}
where $r_{1,2} = I_{1,2} / I_0$ are the normalized critical currents of the lateral junctions.

\begin{figure}[t]
\includegraphics[width=\columnwidth]{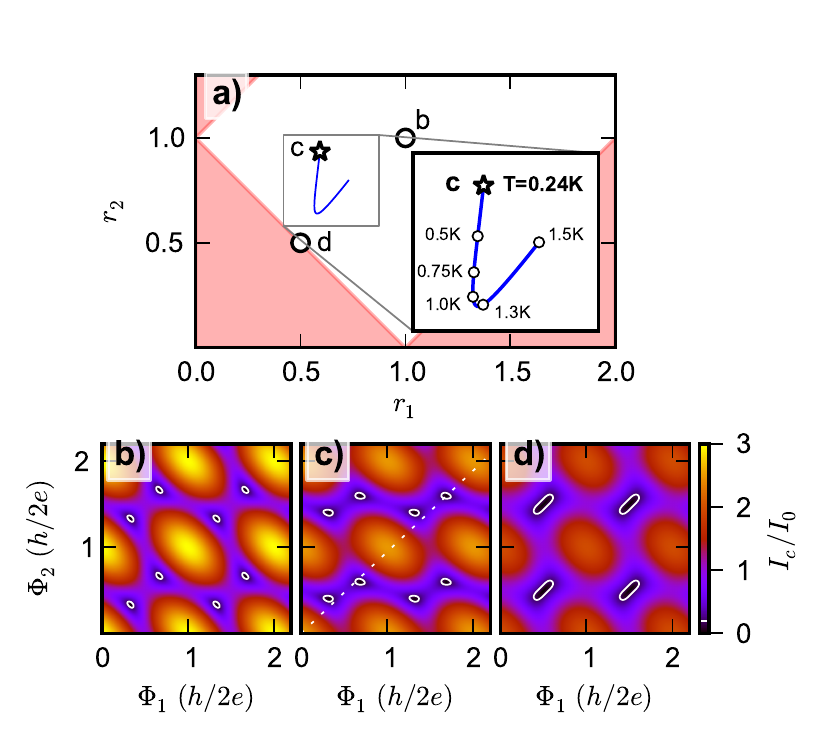}
\caption{
\textbf{a)} Phase diagram for possible realizations of the double-loop interferometer; 
parameters $r_1 = I_1/I_0$ and $r_2 = I_2/I_0$ 
define the amount of asymmetry in critical current between the lateral and the central junctions; a fully symmetric interferometer
is obtained from $r_1 = r_2 = 1$ (b), a partially symmetric interferometer is obtained from $r_1=r_2 \neq 1$
(d); 
the device presented (c) shows an asymmetric configuration at 0.24K, where $r_1 \neq r_2 \neq 1$.
The area shaded in red represents interferometers which are un-balanceable due to excessive asymmetry; the inset shows
the temperature-dependent behaviour of the fabricated device.
\textbf{b, c, d)} Colormap plots of the critical current for the three configurations considered in panel \textbf{a}
as a function of magnetic fluxes linked to each loop ($\Phi_1$, $\Phi_2$); local minima are encircled by the 
white $I_c/I_0 = 0.2$  isoline as a visualization aid; the dotted white line in panel \textbf{c} highlights the flux pair values that
can be set by applying an external omogeneous magnetic field to the slightly asymmetric loops of the presented device.
}
\label{fig:model}
\end{figure}

Examples of interferometers having variable degrees of asymmetry can be appreciated
in the $r_1-r_2$ representation shown in Fig.~\ref{fig:model}a).
The most symmetric case, labeled with the letter b, corresponds to an interferometer in which the three junctions 
have identical critical current values
($r_1 = r_2 = 1$); the corresponding $I_c(\Phi_1, \Phi_2)$ map is shown as a color plot in Fig.~\ref{fig:model}b) and reaches maximum
values $\max(I_c) = 3 I_0$.
A reduced symmetry is represented by the case in which the lateral junctions have identical critical current values, but differ
from the central junction (\textit{e.g.}, $r_1=r_2=0.5$, labeled as d); Figure~\ref{fig:model}d) shows the corresponding $I_c(\Phi_1, \Phi_2)$
map. Finally, the generic asymmetric case is represented by $r_1=0.6$, $r_2=0.9$, values which have been found to approximate 
the behaviour of the presented device at temperature $T=0.24$K; this case is labeled with the letter c, and its 
$I_c(\Phi_1, \Phi_2)$ dependence is presented in Fig.~\ref{fig:model}c). All three cases fulfill Eqns.~\ref{eqn:triangle}, 
showing exact supercurrent suppression for appropriate $\Phi_1$, $\Phi_2$ values.

In our setup, magnetic flux biasing is provided by an external homogeneous magnetic field $B \propto I_{coil}$\footnote{Proportionality
coefficient $\approx 40 \, \mathrm{mT/A}$}, so that the flux values $\Phi_1$ and $\Phi_2$
are proportional to the external field and can differ only as a consequence of asymmetry in the effective area values 
of the superconducting loops:
\begin{equation}
	\Phi_{1,2} = (1 \pm \alpha) A_{eff} B = (1 \pm \alpha) \Phi \, ,
	\label{eqn:field}
\end{equation}
where $A_{eff}$ is the average loop effective area, $\Phi = A_{eff}B$ is the average magnetic flux bias value
and $\alpha$ is the effective area asymmetry coefficient.

\begin{figure}[t]
\includegraphics[width=\columnwidth]{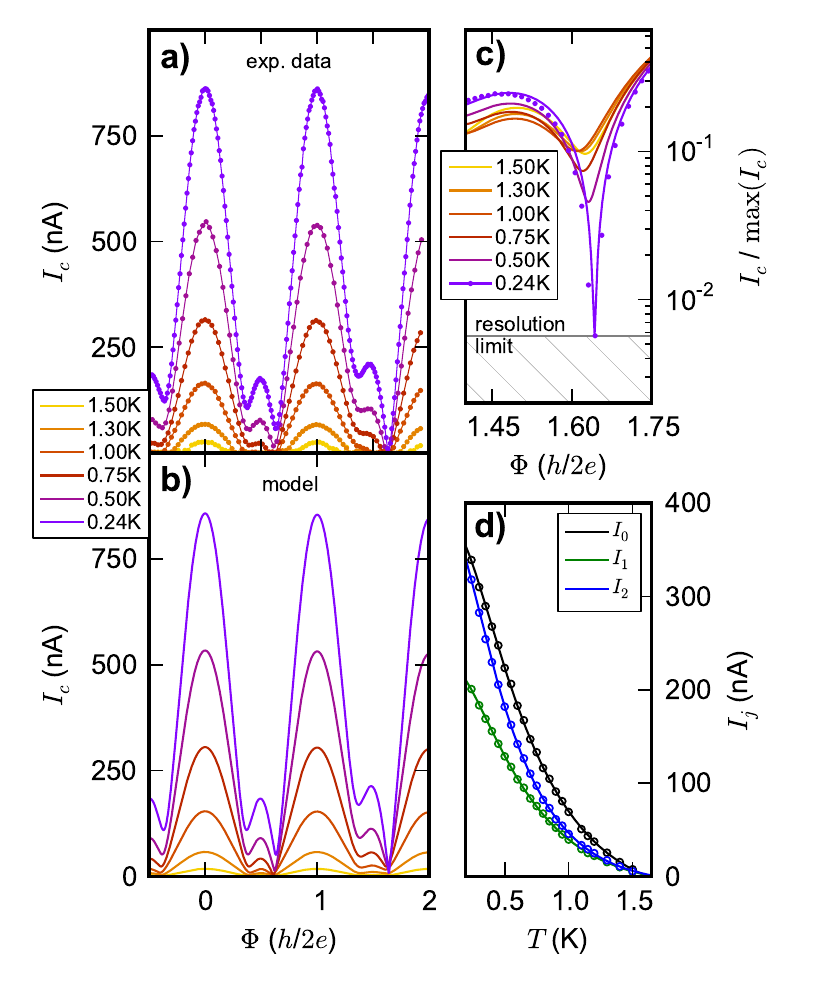}
\caption{
\textbf{a)} Experimental critical current values for the interferometer as a function of magnetic flux at different temperatures (color
coded).  The critical current points have been extracted from $V_{out}(I_{bias}, I_{coil})$ fixed-temperature datasets,
an instance of which has been presented in Fig.~\ref{fig:ivs}a); 
continuous lines joining data points have been added as a visual aid.
\textbf{b)}~Proposed model (Eqns.~\ref{eqn:modelic},~\ref{eqn:field}) fitted to data presented in panel~\textbf{a}. 
\textbf{c)}~Supercurrent suppression \lbrack $I_c(\Phi)$ normalized to the maximum value of $I_c$\rbrack~achieved at $\Phi/\Phi_0 = 5/3$ 
for different values of temperature. Colored continuous lines are derived from the optimal model presented in panel~\textbf{b};
data points calculated for the $T=0.24$K dataset are also displayed, along with a shaded area representing the resolution limit due to 
discretization in current scanning.
\textbf{d)}~Temperature dependence of the critical current for each Josephson junction, extracted from the best fit parameters
(Eqns.~\ref{eqn:modelic},~\ref{eqn:field}) to experimental data. 
The size of the circular markers corresponds to the uncertainty of the parameter estimate. The continuous lines
represent the fitted model for the critical current of long diffusive Josephson junctions in the high-temperature regime 
(Eqn.~\ref{eqn:longj}).
\label{fig:temp}}
\end{figure}

A quantitative analysis of the transport properties of the interferometer has been performed by extracting the $I_c(\Phi)$ values 
by fitting differential resistance data with a sigmoid test function. 
This \textit{a posteriori} approach provides us with switching current data \lbrack shown for selected temperature values 
in Fig.~\ref{fig:temp}a)\rbrack~which are associated with an uncertainty derived from the quadrature propagation of the intrinsic sigmoid width
and the current bias discretization error. At low temperature, the switching is sharp and the relative 
uncertainty of the extraction process is limited by the latter term ($\approx 6\perthousand$); by increasing temperature
the ``intrinsic'' sigmoid width gradually takes over, reaching typical values of tens of nA at $T=1.5\mathrm{K}$.

The extracted $I_c(\Phi, T)$ data have been fitted \lbrack Fig.~\ref{fig:temp}b)\rbrack~using Eqns.~\ref{eqn:modelic},~\ref{eqn:field} as a model, 
estimating a loop surface asymmetry $\alpha = 0.028 \pm 0.003$. A direct comparison between data points and model 
\lbrack Figs.~\ref{fig:temp}a,b)\rbrack~demonstrates the effectiveness of the model in describing our mesoscopic SNS interferometer, 
particularly impressive considering the minimal amount of hypotheses on which it is based.
The fitting procedure provides a quantitative estimate for the temperature dependence of the critical currents 
of the three constituent Josephson junctions, shown in Fig.~\ref{fig:temp}d).
The three junctions show markedly different $I_i(T)$, both in terms of the characteristic temperature scale of the supercurrent
suppression (dependent on the Thouless energy $E_{Th} = \hbar D / L^2$, where $D$ is the diffusion coefficient
and $L$ is the length of the diffusive weak link)
and of the magnitude of the supercurrent (affected both by $E_{Th}$ and the transparency of the SN interface).
A quantitative model\cite{Dubos2001}, whose validity in the high temperature regime ($k_B T \gtrsim E_{Th}$) has been 
experimentally verified for V/Cu/V junctions\cite{Garcia2009}, maps the critical current for a long diffusive SNS weak link to:
\begin{equation}
	I_s(T) = \xi \frac{64 \pi k_B T}{e \mathcal{R}} \sum_{n=0}^\infty
	\frac{\sqrt{ \frac{2 \omega_n}{E_{Th}}} \Delta^2(T)\, \exp \left[ - \sqrt{ \frac{ 2 \omega_n}{E_{Th}}} \,\right]}
	{\left[ \omega_n + \Omega_n + \sqrt{2(\Omega_n^2+\omega_n\Omega_n)}\right]^2} \, ,
	\label{eqn:longj}
\end{equation}
with $\omega_n(T) = (2n+1)\pi k_B T$ and $\Omega_n(T) = \sqrt{ \Delta^2(T) + \omega_n^2(T)}$, where $k_B$ is Boltzmann's constant,
$e$ is the elementary charge, $\mathcal{R} = 3 R_n \approx 30.8\,\Omega$ 
is the average normal state resistance of a single junction of the presented device as extracted from current vs voltage measurements,
 $\xi$ accounts for non-ideality of the normal-superconductor interface  and $\Delta(T)$ is the superconducting gap at the electrodes, 
whose temperature dependence
has been assumed to be BCS-like (parametrically determined by specifying an effective critical temperature $T_c^*$ 
for the vanadium electrodes).
This model has been used to fit $I_i(T)$ data for each junction obtaining the parameter estimates reported in Tab.~\ref{tab:fit}.
Even though the interferometer has been designed to be symmetric, deviations from ideality inherent to the fabrication process
resulted in junctions with quantitatively different $E_{Th}$ and $\xi$ values. 

\begin{table}
 \begin{ruledtabular}
\begin{tabular}{cccc}
$ \, \, $	 &  $\xi$ & $E_{Th}$ ($\upmu$eV) & $T_c^*$ (K) \\
\hline
$I_0$ &  0.211 $\pm$  0.002 &  15.1 $\pm$ 0.2  & 1.61 $\pm$ 0.01\\
$I_1$ &  0.132 $\pm$ 0.003 & 13.7 $\pm$ 0.4 & 1.66 $\pm$ 0.03 \\
$I_2$ &  0.265 $\pm$ 0.004 & 10.2 $\pm$ 0.2 & 1.68 $\pm$ 0.03\\
\end{tabular}
\end{ruledtabular}
\caption{
\label{tab:fit}
Parameter estimates for the ideality coefficient $\xi$, Thouless energy $E_{Th}$ and effective critical temperature at the electrodes $T_c^*$
of the three junctions of the interferometer obtained by fitting Eqn.~\ref{eqn:longj} to $I_j(T)$ data \lbrack Fig.~\ref{fig:temp}d)\rbrack. 
}
\end{table}

The presence of measurably different Thouless
energy scales introduces a temperature dependence in the $r_1-r_2$ parameters for the presented device, as it can be appreciated in
the inset of Fig.~\ref{fig:model}a).
Nevertheless, under optimal flux biasing we were able
to measure supercurrent suppression values lower than $6 \perthousand$ at base temperature \lbrack Fig.~\ref{fig:temp}c)\rbrack, thus 
confirming the fitness of the double-loop geometry as a means to circumvent junction asymmetry in mesoscopic SNS-based devices.

Finally, it is worth noting that the additional degree of freedom granted by the second loop in our geometry entails
the possibility of having the interferometer respond both to the homogeneous part and to the first spatial derivative of the magnetic
field (proportional to the sum and difference of $\Phi_1$ and $\Phi_2$, respectively) on a micrometric length scale; 
moreover, the relative strength of response can be
tuned by designing the interferometer with sensible $r_1-r_2$ parameter values\footnote{For example, selectively gradiometric response
capability is evident in the asymmetric interferometer whose $I_c(\Phi_1,\,\Phi_2)$ is depicted in Fig.~\ref{fig:model}d).},
easily allowed by the flexibility of the shadow-mask lithographic technique.

The authors acknowledge the Italian Ministry of Defense through the PNRM project
``TERASUPER'', the Marie Curie Initial Training Action (ITN) Q-NET 264034 for partial
financial support. 
C.A. thanks the Tuscany Region for funding his fellowship via the 
CNR joint project ``PROXMAG''.
A.R. thanks Fondazione Tronchetti Provera for funding his Ph.D. scholarship
in Scuola Normale Superiore.

\end{document}